\documentclass[9pt]{article}
\usepackage[preprint]{spconf}
\usepackage{amsmath,graphicx,amssymb,comment,lipsum}
\usepackage{multirow,booktabs,array,float,enumerate,dblfloatfix}

\title{STRUCTURE-INFORMED POSITIONAL ENCODING FOR MUSIC GENERATION}
\name{Manvi Agarwal \qquad Changhong Wang \qquad Ga\"{e}l Richard
\thanks{This work was funded by the European Union (ERC, HI-Audio, 101052978). Views and opinions expressed are however those of the author(s) only and do not necessarily reflect those of the European Union or the European Research Council. Neither the European Union nor the granting authority can be held responsible for them. This work was also executed with the help of HPC resources from GENCI–IDRIS (Grant 2022-AD011013986).\\Code, corrected alignments and audio samples can be found on our companion website at \textit{bit.ly/structurepe}}
}
\address{LTCI, T\'{e}l\'{e}com Paris, Institut Polytechnique de Paris, France}
\copyrightnotice{\copyright 2023 IEEE. Personal use of this material is permitted. Permission from IEEE must be obtained for all other uses, in any current or future media, including reprinting/republishing this material for advertising or promotional purposes, creating new collective works, for resale or redistribution to servers or lists, or reuse of any copyrighted component of this work in other works.}

\begin{document}
%
\maketitle
%
\begin{abstract}

Music generated by deep learning methods often suffers from a lack of coherence and long-term organization. Yet, multi-scale hierarchical structure is a distinctive feature of music signals. To leverage this information, we propose a \emph{structure-informed positional encoding} framework for music generation with Transformers. We design three variants in terms of absolute, relative and non-stationary positional information. We comprehensively test them on two symbolic music generation tasks: next-timestep prediction and accompaniment generation. 
As a comparison, we choose multiple baselines from the literature and demonstrate the merits of our methods using several musically-motivated evaluation metrics. 
In particular, our methods improve the melodic and structural consistency of the generated pieces.

\end{abstract}


\section{Introduction}
\label{sec:intro}

Music generation using deep learning is seeing growing interest due to not only its commercial opportunities but also the challenges of recreating creativity with machines. 
Recent improvements in data-driven music generation systems have been influenced by advances in Natural Language Processing (NLP)~\cite{bigo_hdr_2023}, such as the use of Transformers~\cite{vaswani_attention_2017} and tokenization for converting music into symbolic sequences, typically in MIDI or MIDI-inspired formats~\cite{huang_pop_2020,yi_popmag_2020}.
Despite these advances, symbolic music generated by Transformers lacks the rich, multi-scale structures that are a characteristic feature of real music~\cite{ji_survey_2023, wu_jazz_2020}. In addition to their role in the enjoyment of music, such structural qualities can guide music generation by expressing knowledge about the data domain.

Previous works have hinted at the efficacy of incorporating knowledge about musical structure into data-driven models. Symbolic representations, such as REMI~\cite{huang_pop_2020}, compound word tokens~\cite{hsiao_compound_2021} and MuMIDI~\cite{yi_popmag_2020} expand the event vocabulary with tokens for musical structures, such as chord and beat.
However, this obfuscates the representation of syntax or structure with that of semantics or content and causes sequence length and vocabulary size to grow.
Vocabulary choice is a complex issue~\cite{chen_large_2019, zheng_allocating_2021} and increasing vocabulary size compromises the expressivity of multi-layered, deep models~\cite{wies_transformer_2021}.

Instead, many works use the Positional Encoding (PE) module as a candidate for inserting musically-informed priors into Transformers. However, they rely on structural cues that are trivially inferred from the input. PopMAG uses bar and position~\cite{yi_popmag_2020}, SymphonyNet uses note order, measure order and track ID~\cite{liu2022symphony} and RIPO attention uses relative pitch and onset information~\cite{guo_domain_2023}. This leaves space to improve how structural knowledge is encoded and represented by PE.

In this paper, we explore whether providing hierarchical, musically-aware structural information, obtained non-trivially from the input by signal processing methods or human-provided annotations, to the PE module can be beneficial for Transformers for music generation.

\textit{First}, we present a novel, structure-informed positional encoding framework called StructurePE. It comes in three flavours: Structure Absolute Positional Encoding (S-APE), Structure Relative Positional Encoding (S-RPE) and Structure Relative Positional Encoding with a nonstationary kernel (NS-RPE). Evaluating these variants on next-timestep prediction and accompaniment generation, we demonstrate that our methods outperform the baselines on music-focused metrics.

\textit{Second}, to obtain high-quality annotations, we correct the alignment between structural labels and songs in the POP909 dataset~\cite{wang_pop909_2020}, providing this via the companion website.

\textit{Third}, drawing on previous work from NLP, we use Transformers without Positional Encoding (NoPE) as a baseline. We show that, though NoPE is often left out as a baseline for music generation, it is at par with other baselines on both tasks. We argue that NoPE should be included in future work on music generation.


\section{Methods} \label{sec:methods}

\subsection{Input Representation}

We use a binary pianoroll representation for the input, using a resolution of 16 timesteps for one quarter note. A pianoroll is $\mathbf{X} \in \mathbb{B}^{(n_\text{tracks} \times 128) \times n_\text{time}}, \mathbb{B} = \{0, 1\}$. 
Every pianoroll column $\mathbf{x}_t, t \in \{1, ..., n_\text{time}\}$, is associated with multiple levels of structural labels that describe the broader context of the pitches of that column. For example, $\mathbf{x}_t$ might be part of a melody, a chord and a phrase.

\subsection{Positional Encoding} \label{sec:pe}

\begin{figure}[t]
    \centering
    \includegraphics[width=\columnwidth]{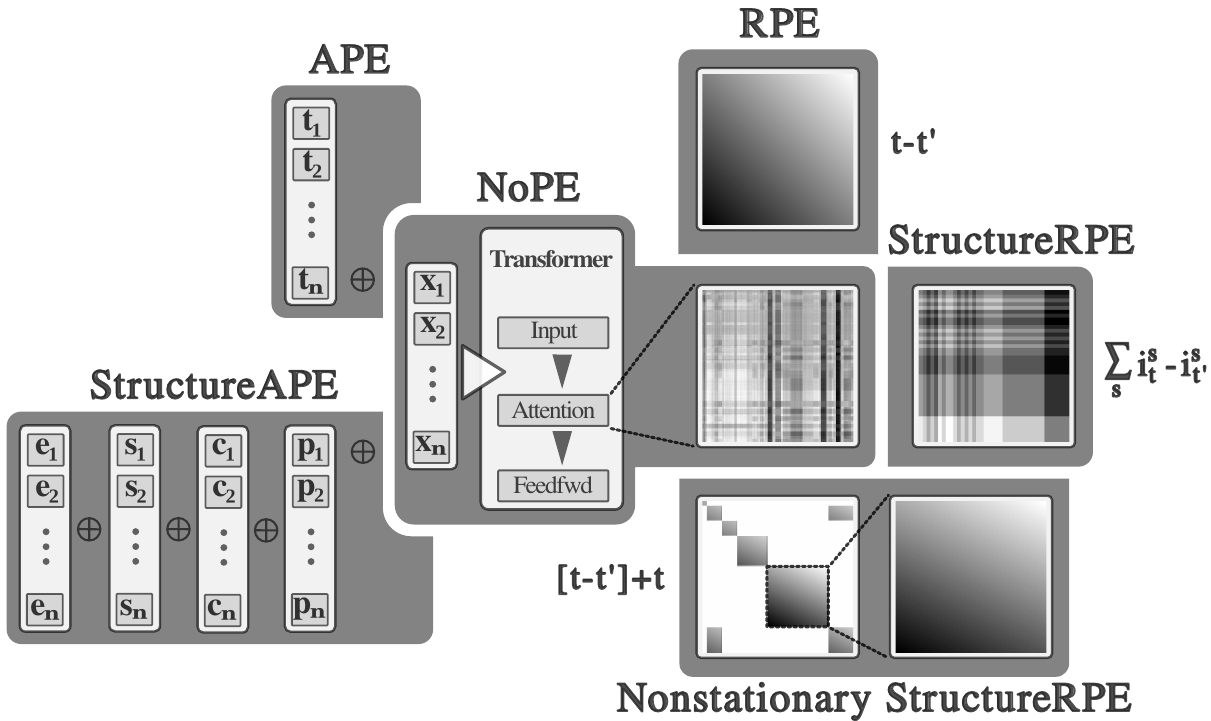}
    \caption{Illustrative schematic of PEs - both baselines (NoPE, APE and RPE) and ours (rest) - and their use in Transformers. See Sections \ref{sec:pe} and \ref{sec:structpe} for details.}
    \label{fig:summary}
\end{figure}

Without PE, Transformers are invariant to the order of the input~\cite{vaswani_attention_2017}. PE gives the model information about what content in the sequence occurs in what position. Positional information is injected into the model with two techniques: Absolute Positional Encoding (APE) and Relative Positional Encoding (RPE). As shown in Fig. \ref{fig:summary}, in APE, positional information is added to the input before it enters the Transformer. Each timestep $t$ in the sequence is mapped to a positional index $i_t$, which is embedded as a vector using $p_{\text{APE}}(\cdot)$. Typically, the former is $i_t = f(t) = t$ and the latter is a sinusoidal embedding function~\cite{vaswani_attention_2017}. In RPE, positional information is injected directly into the attention matrix. Each pair of timesteps $(t,t^{\prime})$ is mapped to a value that represents the relationship between them. Usually, RPE uses $f(t, t^{\prime}) = t - t^{\prime}$. Similar to APE, RPE uses an embedding function $p_{\text{RPE}}(\cdot)$ on $f(t,t^{\prime})$~\cite{shaw_relative_2018}. 

\subsection{Structure-informed Positional Encoding} \label{sec:structpe}

For StructurePE, we map each timestep $t$ to its structural vector $\textbf{i}_t = [ i_t^s ] $ where $s \in S = \{ \texttt{tempo}, \texttt{ section}, \texttt{ chord},\\\texttt{mpitch} \}$. 
This gives us a multi-dimensional, hierarchical positional vector, each of whose elements correspond to different aspects of musical structure. We propose three variants of StructurePE.

\vspace{0.1cm}
\noindent \textbf{StructureAPE (S-APE)} \hspace{0.05cm} Following the APE method~\cite{vaswani_attention_2017}, we use an embedding function to map each $i_t^s$ to a vector and add these vectors to the input:

\begin{equation}
    \text{input}_t = \textbf{x}_t + \sum_{s \in S} p_{\text{APE}}(i_t^s)
\end{equation}
We experimented with two embedding functions $p_{\text{APE}}(\cdot)$ : a learnable embedding (L S-APE), and a sinusoidal embedding (S S-APE)~\cite{guo_domain_2023}.

\vspace{0.1cm}
\noindent \textbf{StructureRPE (S-RPE)} \hspace{0.05cm} We take the RPE approach~\cite{shaw_relative_2018} and use the relative distance between $i_t^s$ and $i_{t^{\prime}}^s$ for all timesteps $t, t^{\prime}$. The distances are embedded as vectors and incorporated directly into the attention matrix as follows:

\begin{equation} \label{eq:SRPE}
    z_{t, t^{\prime}} = \textbf{q}_{t} \textbf{k}_{t^{\prime}}^T + \sum_{s \in S} \textbf{q}_{t} p_{\text{RPE}} (i_t^s - i_{t^{\prime}}^s)^T
\end{equation}
where $\textbf{q}_{t}$ is the query and $\textbf{k}_{t^{\prime}}$ is the key.
Similar to S-APE, we have two choices for the embedding function $p_{\text{RPE}}(\cdot)$, giving us L S-RPE and S S-RPE.

\vspace{0.1cm}
\noindent \textbf{Nonstationary StructureRPE (NS-RPE)} \hspace{0.05cm} Following a line of work that identifies the attention mechanism as a kernel~\cite{tay_survey_2022}, the standard RPE formulation, which is also used by S-RPE, can be interpreted as implementing a stationary kernel~\cite{liutkus_relative_2021, choromanski_learning_2023}. S-RPE is stationary with respect to both time $t$ and the structural positional indices $i_t^s$. 

While the kernel formulation was introduced to reduce the complexity of the attention calculation, we can gain representational power by considering the broader category of nonstationary kernels. Nonstationary kernels can express rich relationships between positions and do not depend solely on the lag between them~\cite{genton_classes_2001}. We add a kernel $\kappa_{\varphi} ( t, t^{\prime} )$ to Eq. \ref{eq:SRPE} to introduce nonstationarity with respect to time at a chosen structural level $\varphi \in S$. $\kappa_{\varphi} ( t, t^{\prime} )$ gives the model information about \textit{where} in the sequence the lag $t - t^{\prime}$ occurs, creating an input-dependent kernel that is not invariant to translation.
    \begin{equation} \label{eq:ns_1}
        z_{t, t^{\prime}} = \textbf{q}_{t} \textbf{k}_{t^{\prime}}^T + \sum_{s \in S} \textbf{q}_{t} p_{\text{RPE}} (i_t^s - i_{t^{\prime}}^s)^T + \textbf{q}_{t} \kappa_{\varphi} (t, t^{\prime})^T
    \end{equation}

    \begin{equation} \label{eq:ns_2}
        \kappa_{\varphi} (t, t^{\prime}) = \begin{cases}
                                    p_{\text{RPE}} (t - t^{\prime}) + p_{\text{RPE}} (t) &\text{if $i_t^{\varphi} = i_{t^{\prime}}^{\varphi}$}\\
                                    0 &\text{otherwise}
                                \end{cases} 
    \end{equation}

To reduce computation, we only consider those pairs $(t, t^{\prime})$ which have the same structural label within the category $\varphi$ (e.g., they are labelled with the same chord or same phrase). This choice is informed by our intuition that temporal relationships within the same structure should remain identical regardless of where in the sequence the structure occurs.

\begin{figure*}[b]
    \centering
    \includegraphics[width=2.0\columnwidth]{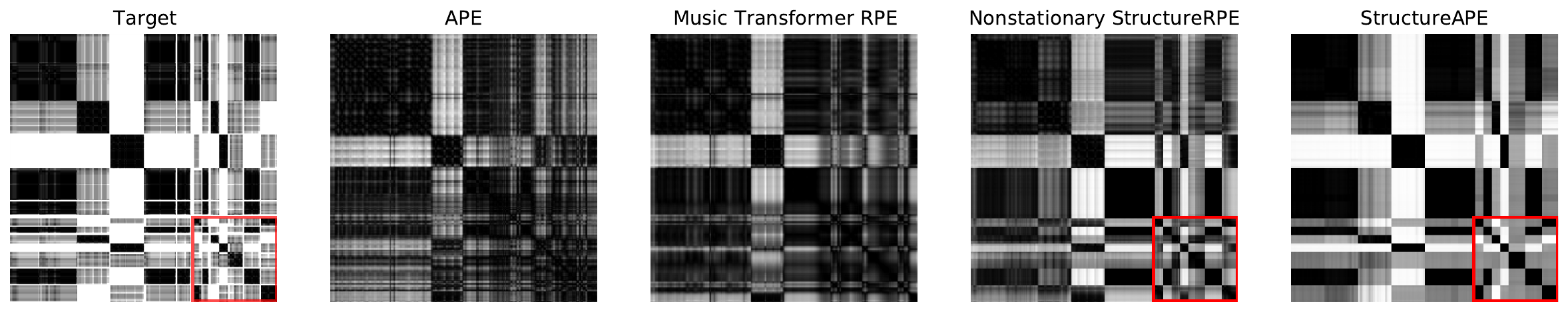}
    \caption{Comparison of self-similarity matrices from music generated by baselines (columns 2 and 3) and our methods (columns 4 and 5). Best viewed in colour.}
    \label{fig:ssm_full}
\end{figure*}

\section{Experiments} \label{sec:experiments}

\subsection{Task setup} \label{sec:task}
We select two conditional music generation tasks:

\vspace{0.1cm}
\noindent \textbf{Next-timestep prediction} \hspace{0.05cm} Given input $\{ \mathbf{x}_1, ... , \mathbf{x}_L \}$, where $\mathbf{x}_t \in \mathbb{B}^{(n_\text{tracks} \times 128)}$ is a column of the pianoroll for timestep $t$, produce as output $\mathbf{x}_{L+1}$.

\vspace{0.1cm}
\noindent \textbf{Accompaniment generation} \hspace{0.05cm} Given input $\{ \mathbf{x}_1, ... , \mathbf{x}_L \}$, where $\mathbf{x}_t \in \mathbb{B}^{((n_\text{tracks} - 1) \times 128)}$ gives the melody and bridge tracks, generate $\{ \mathbf{y}_1, ... , \mathbf{y}_L \}$, where $\mathbf{y}_i \in \mathbb{B}^{128}$ gives the piano track.

We vary the training length $L_1$ and testing length $L_2$ for varying levels of difficulty, parameterized as $(L_1, L_2)$. For next-timestep prediction, we use settings \textbf{N1}:$(512, 1024)$ and \textbf{N2}:$(1024,1024)$. For accompaniment generation, we explore settings \textbf{A1}:$(512,512)$, \textbf{A2}:$(512, 1024)$, and \textbf{A3}:$(1024,1024)$. 

\subsection{Model and Dataset} \label{sec:model}

We use a 2-layer Transformer decoder with 4 heads, trained with a learning rate scheduler, early stopping and curriculum learning~\cite{bengio_curriculum_2009} on the sequence length.
We use the Chinese POP909 dataset~\cite{wang_pop909_2020,dai_automatic_2020}. 
Every song contains four labels, at different temporal resolutions, at each timestep - tempo, section, chord and melody - but they are not aligned with the songs due to two variable-length errors: (i) silence at the start, and (ii) musical phrases that are not part of the labeling, such as pickup measures. Although (i) is trivial to eliminate, (ii) cannot be corrected with high precision in an automated way. So, we manually select the beat position at which the structural labels should start and provide this as a supplement to the dataset on our companion website. Further details on the dataset, training procedure and data alignment method are also provided there.

\subsection{Baselines}

We consider two types of baselines: (i) general PE, and (ii) PE designed with structure for music generation. For (i), we have Transformers without PE (NoPE), with APE~\cite{vaswani_attention_2017} and with RPE~\cite{shaw_relative_2018,huang_music_2018}, as presented in Section \ref{sec:pe}. For (ii), we consider SymphonyNet~\cite{liu2022symphony} as the baseline for structurally-informed APE (S-APE/b) and use note order and measure order for structural labels. We also consider RIPO attention~\cite{guo_domain_2023} as the baseline for structurally-informed RPE (S-RPE/b) and use relative pitch and onset information as relative structural distances. 
We consider two possibilities for $\varphi$ in NS-RPE: $\varphi = \texttt{chord}$ (NS-RPE/c) and $\varphi = \texttt{section}$ (NS-RPE/s). Our models differ from the baselines only in the PE method being used.

\begin{table*}[t]
\centering
\resizebox{2.05\columnwidth}{!}{%
\begin{tabular}{ccccccccccccccccccccc}
\toprule
\multicolumn{1}{c|}{\multirow{4}{*}{\textbf{Methods}}} & \multicolumn{8}{c|}{\textbf{Next-timestep Prediction}} & \multicolumn{12}{c}{\textbf{Accompaniment Generation}} \\ 
\multicolumn{1}{c|}{} & \multicolumn{4}{c}{\textbf{N1}} & \multicolumn{4}{c|}{\textbf{N2}} & \multicolumn{4}{c}{\textbf{A1}} & \multicolumn{4}{c}{\textbf{A2}} & \multicolumn{4}{c}{\textbf{A3}} \\ 
\multicolumn{1}{c|}{} & \multicolumn{1}{c}{\textbf{SSMD}} & \multicolumn{1}{c}{\textbf{CS}} & \multicolumn{1}{c}{\textbf{GS}} & \multicolumn{1}{c|}{\textbf{NDD}} & \multicolumn{1}{c}{\textbf{SSMD}} & \multicolumn{1}{c}{\textbf{CS}} & \multicolumn{1}{c}{\textbf{GS}} & \multicolumn{1}{c|}{\textbf{NDD}} & \multicolumn{1}{c}{\textbf{SSMD}} & \multicolumn{1}{c}{\textbf{CS}} & \multicolumn{1}{c}{\textbf{GS}} & \multicolumn{1}{c|}{\textbf{NDD}} & \multicolumn{1}{c}{\textbf{SSMD}} & \multicolumn{1}{c}{\textbf{CS}} & \multicolumn{1}{c}{\textbf{GS}} & \multicolumn{1}{c|}{\textbf{NDD}} & \multicolumn{1}{c}{\textbf{SSMD}} & \multicolumn{1}{c}{\textbf{CS}} & \multicolumn{1}{c}{\textbf{GS}} & \textbf{NDD} \\ 
\multicolumn{1}{c|}{} & \multicolumn{1}{c}{$\downarrow$} & \multicolumn{1}{c}{$\uparrow$} & \multicolumn{1}{c}{$\uparrow$} & \multicolumn{1}{c|}{$\downarrow$} & \multicolumn{1}{c}{$\downarrow$} & \multicolumn{1}{c}{$\uparrow$} & \multicolumn{1}{c}{$\uparrow$} & \multicolumn{1}{c|}{$\downarrow$} & \multicolumn{1}{c}{$\downarrow$} & \multicolumn{1}{c}{$\uparrow$} & \multicolumn{1}{c}{$\uparrow$} & \multicolumn{1}{c|}{$\downarrow$} & \multicolumn{1}{c}{$\downarrow$} & \multicolumn{1}{c}{$\uparrow$} & \multicolumn{1}{c}{$\uparrow$} & \multicolumn{1}{c|}{$\downarrow$} & \multicolumn{1}{c}{$\downarrow$} & \multicolumn{1}{c}{$\uparrow$} & \multicolumn{1}{c}{$\uparrow$} & $\downarrow$ \\ \hline \hline
\multicolumn{21}{c}{\textbf{Baselines}} \\ \hline \hline
\multicolumn{1}{c|}{\textbf{NoPE}} & 7.20 & \textbf{95.13} & 94.19 & \multicolumn{1}{c|}{13.58} & 7.21 & 95.12 & 93.52 & \multicolumn{1}{c|}{13.59} & 53.09 & 65.57 & 33.55 & \multicolumn{1}{c|}{44.32} & 54.85 & 65.15 & 69.61 & \multicolumn{1}{c|}{44.16} & 54.99 & 64.83 & 69.33 & 46.31 \\ 
\multicolumn{1}{c|}{\textbf{APE}} & 10.62 & 88.74 & 91.35 & \multicolumn{1}{c|}{21.45} & 6.92 & 93.65 & 90.98 & \multicolumn{1}{c|}{12.73} & 49.48 & 68.27 & 22.77 & \multicolumn{1}{c|}{43.77} & 54.42 & 66.08 & 48.23 & \multicolumn{1}{c|}{43.90} & 49.87 & 68.42 & 45.13 & 43.57 \\ 
\multicolumn{1}{c|}{\textbf{RPE}} & 6.77 & 93.62 & 92.70 & \multicolumn{1}{c|}{12.75} & \textbf{6.79} & 93.66 & 90.27 & \multicolumn{1}{c|}{\textbf{12.51}} & 49.46 & 66.88 & 30.43 & \multicolumn{1}{c|}{48.64} & 50.61 & 66.98 & 65.16 & \multicolumn{1}{c|}{55.56} & 49.26 & 68.26 & 48.10 & 61.42 \\ \hline 
\multicolumn{1}{c|}{\textbf{S-APE/b}} & 7.21 & \textbf{95.13} & 93.44 & \multicolumn{1}{c|}{13.58} & 7.21 & 95.11 & 94.08 & \multicolumn{1}{c|}{13.59} & 52.99 & 65.48 & 33.54 & \multicolumn{1}{c|}{44.14} & 54.63 & 64.99 & 69.74 & \multicolumn{1}{c|}{44.10} & 55.07 & 64.79 & 69.38 & 46.96 \\ 
\multicolumn{1}{c|}{\textbf{S-RPE/b}} & 7.21 & \textbf{95.13} & 93.84 & \multicolumn{1}{c|}{13.60} & 7.20 & \textbf{95.14} & 93.83 & \multicolumn{1}{c|}{13.59} & 53.04 & 65.46 & 33.71 & \multicolumn{1}{c|}{52.84} & 54.70 & 65.02 & 69.58 & \multicolumn{1}{c|}{62.26} & 54.74 & 64.74 & 69.27 & 77.72 \\ \hline \hline
\multicolumn{21}{c}{\textbf{Our Methods}} \\ \hline \hline
\multicolumn{1}{c|}{\textbf{L S-APE}} & 7.42 & 94.82 & 94.41 & \multicolumn{1}{c|}{14.17} & 7.46 & 94.74 & 93.24 & \multicolumn{1}{c|}{14.25} & \textbf{30.65} & 73.74 & \textbf{34.41} & \multicolumn{1}{c|}{43.92} & \textbf{30.47} & 73.78 & 70.52 & \multicolumn{1}{c|}{43.66} & \textbf{30.22} & 73.81 & \textbf{71.42} & 43.94 \\ 
\multicolumn{1}{c|}{\textbf{S S-APE}} & 7.24 & 95.03 & 93.62 & \multicolumn{1}{c|}{13.71} & 7.27 & 95.05 & 92.96 & \multicolumn{1}{c|}{13.74} & 31.14 & \textbf{75.20} & 34.21 & \multicolumn{1}{c|}{\textbf{43.62}} & 31.16 & \textbf{75.19} & \textbf{70.79} & \multicolumn{1}{c|}{\textbf{43.52}} & 31.99 & \textbf{74.77} & 70.60 & \textbf{43.56} \\ \hline 
\multicolumn{1}{c|}{\textbf{L S-RPE}} & 7.20 & 95.05 & \textbf{94.84} & \multicolumn{1}{c|}{13.58} & 7.19 & 95.07 & 94.62 & \multicolumn{1}{c|}{13.59} & 39.27 & 69.73 & 32.82 & \multicolumn{1}{c|}{44.60} & 38.23 & 70.82 & 68.92 & \multicolumn{1}{c|}{45.25} & 37.56 & 70.73 & 69.30 & 45.74 \\ 
\multicolumn{1}{c|}{\textbf{S S-RPE}} & 7.16 & \textbf{95.13} & 94.16 & \multicolumn{1}{c|}{13.53} & 7.18 & 95.07 & \textbf{94.77} & \multicolumn{1}{c|}{13.57} & 40.37 & 68.88 & 33.52 & \multicolumn{1}{c|}{44.61} & 39.30 & 70.13 & 70.05 & \multicolumn{1}{c|}{45.68} & 38.54 & 69.65 & 69.99 & 47.16 \\ \hline 
\multicolumn{1}{c|}{\textbf{NS-RPE/c}} & 6.84 & 93.76 & 93.17 & \multicolumn{1}{c|}{12.84} & 6.85 & 93.83 & 93.19 & \multicolumn{1}{c|}{12.77} & 39.99 & 69.99 & 33.22 & \multicolumn{1}{c|}{46.37} & 39.17 & 71.24 & 69.17 & \multicolumn{1}{c|}{47.43} & 39.64 & 70.88 & 69.44 & 47.63 \\ 
\multicolumn{1}{c|}{\textbf{NS-RPE/s}} & \textbf{6.76} & 93.77 & 90.61 & \multicolumn{1}{c|}{\textbf{12.62}} & $\diagdown$ & $\diagdown$ & $\diagdown$ & \multicolumn{1}{c|}{$\diagdown$} & 40.26 & 69.19 & 33.59 & \multicolumn{1}{c|}{50.59} & 39.38 & 70.36 & 70.10 & \multicolumn{1}{c|}{47.25} & $\diagdown$ & $\diagdown$ & $\diagdown$ & $\diagdown$ \\ \bottomrule
\end{tabular}%
}
\caption{Results for all tasks (see Section \ref{sec:task}) and all metrics (see Section \ref{sec:metrics}). SSMD: self-similarity matrix distance, CS: chroma similarity, GS: grooving similarity, NDD: note density distance. $\uparrow$: higher values are better. $\downarrow$: lower values are better.}
\label{tab:results}
\end{table*}

\subsection{Post-processing: Binarization and Velocity-encoding} \label{sec:post_processing}

To compare with the targets, we convert the probability output of the model into binary pianorolls. We adopt four methods of binarization for next-note prediction and use the mean square error to select the best method: (1) Threshold: use a fixed threshold as a step function to binarize. (2) Threshold with merge: after applying (1), fill the gaps between two occurrences of the same pitch when the gap length is below a certain value. 
(3) Top-$k$ sampling: select the $k$ highest probabilities and sample randomly from these.
(4) Top-$k$ sampling with merge: do (3), then use merge technique from (2).
For accompaniment generation, we employ a velocity-encoding method to allow continuous-valued pianorolls with expressive dynamics. For this, we linearly map the probability output to loudness.

\subsection{Evaluation metrics}\label{sec:metrics}
To assess the musical quality of the generated pianorolls, we use metrics from the literature to compare the structure, melody and rhythm of the target and the prediction. 

\vspace{0.1cm}
\noindent \textbf{Self-similarity matrix distance (SSMD)} \hspace{0.05cm} With the chroma profile (number of occurrences of each pitch-class $C, C^{\#}, ... ,\\B$) over time for each pianoroll, we calculate the self-similarity matrix (SSM) as the cosine similarity between pairs of chroma profiles. SSMD is the mean absolute difference between the SSMs of the target and the prediction~\cite{wu_musemorphose_2021}.

\vspace{0.1cm}
\noindent \textbf{Chroma similarity (CS)} \hspace{0.05cm} The chroma onset vector of a half-measure in a pianoroll gives the number of onsets for each pitch-class in that half-measure. CS is given by the cosine similarity of chroma onset vector pairs, where one vector comes from the target pianoroll and the other comes from the corresponding half-measure of the predicted pianoroll~\cite{wu_musemorphose_2021}.

\vspace{0.1cm}
\noindent \textbf{Grooving similarity (GS)} \hspace{0.05cm} For a pianoroll, we obtain a histogram of the number of onsets, calculated at a resolution of $1/16^{th}$  notes. GS is the overlap between the note onset histograms of the target and the prediction. 

\vspace{0.1cm}
\noindent \textbf{Note density distance (NDD)} \hspace{0.05cm} Note density is the number of pitches in every $1/16^{th}$ note. NDD is the mean absolute difference between corresponding $1/16^{th}$ notes of the target and the prediction~\cite{haki_real_2022}.

\section{Results and Discussion}\label{sec:results}

We present the highlights of our findings, which are given in Table \ref{tab:results}. The differences between the PE methods in next-timestep prediction are small. This corresponds well with our comparison of the target and generated pianorolls, which look nearly identical and sound highly similar. In contrast, the results on accompaniment generation are more varied and merit a detailed look.

We begin with SSMD since our focus is the infusion of structural information. We observe that each of our PEs outperform all baselines on SSMD in accompaniment generation, with S-APE achieving the best SSMD by a large margin. For a qualitative perspective on this finding, we plot the SSMs for one sample which falls within the 50 best SSMDs for four different PE methods on setting A2 in Fig. \ref{fig:ssm_full}. The baseline SSMs lack even the coarse structure, while our PEs ably reproduce the large-scale similarity features. On zooming into the bottom-right section of the SSMs, outlined in red in Fig. \ref{fig:ssm_full}, we see that S-APE misses the finer structural details. On the other hand, NS-RPE/c can capture small-scale, high-frequency information. We hypothesize that it is capable of doing so because the nonstationary kernel $\kappa_\varphi$ from Eq. \ref{eq:ns_2} models variation within uniform musical blocks. We can potentially use this to enhance the diversity of generation and produce heterogeneous structures at multiple scales.

Coming to CS, our PEs also comfortably outperform the baselines. While this result underlines the ability of our framework to model melodic features, it also indicates that our structural labels are favourable for incentivizing musically-relevant features, such as tonal consistency. The latter point is made stronger when we compare our StructurePE variants against the baseline structure-informed PEs across tasks and metrics. The better performance of our PEs demonstrates that the type and quality of structural information is crucial.

On GS and NDD, our methods outperform the baselines but the margins are not large. This matches our observation that the generated music, both in the baselines and our methods, are indeed missing notes. However, on listening to the music generated by our models, we argue that it works better as an accompaniment, compared to the baselines. In future work, we will confirm this with a listening study.

Zooming out, we see that NoPE is at par with the other methods on most metrics for both tasks. This surprising finding agrees with previous work from NLP showing that NoPE implicitly and flexibly captures positional information~\cite{tsai_transformer_2019,haviv_transformer_2022}. However, recent work on PE modules for music generation fail to include NoPE as a baseline~\cite{liu2022symphony, guo_domain_2023, hsiao_compound_2021}. We argue that NoPE should be considered a serious contender and included in future work on music generation with Transformers.

Finally, in terms of length generalization, APE performs poorly on N1, which is in line with the literature~\cite{liutkus_relative_2021}. However, it performs at par (CS, SSMD) or better (NDD) than other baselines on A2. This challenges the prevailing notion that, compared to RPE, APE is bad at length generalization as it solely represents absolute positions~\cite{kazemnejad_impact_2023}.

\section{Conclusion}
\label{sec:conclusion}

In this paper, we examined how structural information about musical signals can be leveraged via PE to improve music generation with Transformers. Using a novel framework with three PE variants, we found that, compared to several baselines, our methods particularly improve the structural and melodic properties of the generated music, while also boosting rhythm and polyphony. A qualitative analysis showed that our methods reproduce the overall musical structure of the target and have the ability to represent fine structural details.

\bibliographystyle{IEEEbib}
\bibliography{strings}

\end{document}